\documentstyle[pre,aps]{revtex}

\renewcommand{\S}{\hat K}
\newcommand \r {\rho}
\newcommand \la {\lambda}
\newcommand \pr {\prime}
\newcommand \be {\begin{equation}}
\newcommand \ee {\end{equation}}

\begin{document}

\draft

\setcounter{page}{1}
\title{ Strengthened Lindblad inequality: applications
in non equilibrium thermodynamics and
quantum information theory.}

\author{A.E. Allahverdyan, D.B. Saakian }
\address{ Yerevan Physics Institute \\
Alikhanian Brothers St.2, Yerevan 375036, Armenia}
\maketitle
\begin{abstract}
A strengthened Lindblad inequality has been proved. We have applied this
result for proving a generalized $H$-theorem 
in non equilibrium thermodynamics. 
Information processing also 
can be considered as some thermodynamic process. From this point of view
we have proved a strengthened data processing inequality in quantum
information theory.
\end{abstract}

\vspace{15mm}

There are close connections between statistical thermodynamics and 
information
theory \cite{stratinfo}\cite{pop}. It is well known that physical ideas 
played an important role
as sources of information theory \cite{stratinfo}.
On the other hand, the concept of information is crucial 
for understanding some important 
physical problems such as  Maxwells "demon" \cite{pop} or the 
general problem of quantum correlation between two 
subsystems  \cite{barnet}.

In this paper we concentrate on the two connected problems.
These are the $H$-theorem problem  
in non equilibrium quantum statistical thermodynamics 
and the problem of quantum data processing in quantum information theory. 
The concepts of entropy (or other entropy-like
measures) and H-theorem are particularly important in {\it quantum} 
statistical physics because a correct definition is only possible
in the framework of quantum mechanics. In classical theory entropy can 
only be introduced in a somewhat limited and artificial manner
\cite{landau}\cite{wehrl}.

Suppose that a quantum system is 
described by a density matrix $\rho (t)$
at the moment $t$.
In the general case evolution of the non equilibrium system 
in {\it markoffian} regime is described by some
general quantum evolution operator 
\be
\label{1}
\S (t^{\pr },t)\r (t)=\r (t^{\pr })
\ee
In the most general case $\S $ must be linear, completely positive and 
trace-preserving
\cite{shum1}\cite{krauss}, and
has standard representation
\begin{equation}
\label{st3}
\S \rho =\sum_\mu A_\mu ^{\dag }\rho A_\mu ,\ \ \sum_\mu A_\mu
A_\mu ^{\dag }=\hat 1.
\end{equation}
which was introduced in \cite{krauss}. This representation is equivalent 
to the so called unitary representation where 
the non-unitary evolution of the system is regarded as a part of the 
unitary evolution of some larger system. 
Eq. (\ref{st3}) contains unitary transformations,
nonselective measurements, partial traces, et all. For a non-markoffian 
case $\S$ also depends on the "history"
from some initial time $t_0$ to $t^{\prime }$ \cite{stratthermo}.

If the evolution of the system is in the 
{\it stationary markoffian} regime 
then 
\be
\label{2}
\S (t^{\pr },t)
=\S (t^{\pr }-t)
\ee
Stationary markoffian regime is a reasonable conjecture if the system is 
not far from
equilibrium \cite{shlogl}\cite{stratthermo}, or if it can be described by 
a non-hermitian time-independent hamiltonian
\cite{stratthermo} or by  quantum Langevin equations \cite{lax}.

One of the most important quantities which can be defined for statistical 
systems is entropy 
\cite{shlogl}\cite{stratthermo}\cite{stratinfo}\cite{pop}\cite{landau}\cite{wehrl}.
This quantity was introduced in quantum statistical physics
by J. von Neumann.
\be
\label{3}
S(\r )=-{\rm tr}\r \ln \r
\ee
The concept of entropy has at least three main ingredients.
In the first, entropy of a macroscopic statistical system can only 
increase if the system tends to an equilibrium.
Therefore entropy is maximal at this state. This is the well known 
H-theorem. We want to stress that increasing of (\ref{3})
with time can be violated if the evolution of the system is not 
stationary or markoffian \cite{shlogl}\cite{stratthermo}
or the system is open or mesoscopic \cite{open}. 
For example, entropy  can exhibit exactly 
periodic behavior for some open system\cite{knight}. This means that 
(\ref{3}) is not the relevant statistical function for 
such systems.
Entropy is additive function, and also invariant of an unitary 
transformation.

In the second, entropy can be regarded as a measure of the lack of 
information about a system. Therefore $S(\r )$
should increase after coarse-graining procedure 
\cite{stratinfo}\cite{stratthermo}\cite{wehrl}
\begin{equation}
\label{dop1}
S(\sum_i p_i\r _i)\geq \sum_ip_i S(\r _i),\  \ \sum_i p_i=1, \  \ p_i 
\geq 0
\end{equation}
Where the information about coordinate $i$ is lost ($i$ can be also 
continuous).

In the third entropy can be considered as a measure of the amount of 
chaos, or, to what extent the density matrix $\r $
can be considered as "mixed". Indeed, the non negative $S(\r )$ is zero 
for a pure density matrix, and is maximal for
homogeneous $\r $.  Eq (\ref{3}) can also be viewed as one of the basic 
statements of equilibrium
statistical physics \cite{shlogl}\cite{stratthermo}\cite{landau}. For 
example, after several assumptions the most important relation in
thermostatics: $TdS=dE+pdV$ can be derived from (\ref{3}) (where all 
symbols have their ordinary meaning).

Now the following questions arise. Is it possible to define an 
entropy-like function for a mesoscopic statistical system
or for  an open system? Is it possible to save in this definition the 
main aspects of usual entropy? Large number of papers 
and books are devoted to these questions 
\cite{shlogl}\cite{stratthermo}\cite{77}. The answer is {\it "Yes"} 
at least in the case when the evolution
of a system is stationary markoffian, and has a well defined stationary 
distribution. The concrete form of this 
distribution is not important. Before the definition we need some 
mathematics.

Quantum relative entropy between two
density matrices $\rho _1$, $\rho _2$ is defined as follows 
\begin{equation}
\label{st9}
S(\rho _1||\rho _2)={\rm tr}(\rho _1\log \rho _1-\rho _1\log \rho _2).
\end{equation}
This positive quantity was introduced by Umegaki \cite{umegaki}
and characterizes the degree of
'closeness' of density matrices $\rho _1$, $\rho _2$. The properties of
quantum relative information were reviewed by M.Ohya \cite{ohya}. 
Here
only two basic properties  are mentioned.
\begin{equation}
\label{st10}
S(\rho _1||\rho _2)\geq S(\S \rho _1||\S \rho _2).
\end{equation}
\begin{equation}
\label{st11}
S(\la \rho _1+(1-\la )\rho _2||\la \sigma _1+(1-\la )\sigma _2)
\leq \la S(\rho _1||\sigma
_1)+(1-\la )S(\rho _2||\sigma _2).
\end{equation}
Where $0\leq \la \leq 1$. The first inequality was proved 
by Lindblad \cite{lindblad}. 

Now for a system with stationary distribution $\r _{st}$, and markoffian 
stationary evolution operator $\S$ the following
function is defined
\be
\label{5}
-S(\r (t)||\r _{st})
\ee
This function is additive, and also increases after coarse-graining 
procedure as we see from (\ref{st11}). 
Further, eq. (\ref{st10}) which can be written as 
\begin{equation}
\label{dop3}
-S(\S \r (t)|| \r _{st})\geq -S(\r (t)||\r _{st})
\end{equation}
is H-theorem for (\ref{5}).

The definition (\ref{5}) is closely related to the functions which are 
used in usual equilibrium statistical physics.
A very large closed statistical system can be described by microcanonical 
distribution where $\r _{st}$ can be represented as 
an unit matrix (up to some unessential factors).
In this case (\ref{5}) reduces to eq. (\ref{3}) (at least in the case of 
finite dimensional Hilbert space), 
and from (\ref{dop3}) we have the usual H-theorem. Further it is well 
known that 
for a closed macroscopic system canonical and microcanonical 
distributions are equivalent (except some special cases like 
second-order phase transitions). But in some sense canonical distribution 
has larger area of application because it can 
describe some mesoscopic or quasi-open systems 
\cite{stratthermo}\cite{landau}. If we take 
$\r _{st}=\exp (-\beta H)/Z$ in eq. (\ref{5})
(where $\beta $ is  inverse temperature, and $H$ is  hamiltonian) then
\begin{equation}
\label{dop5}
S(\r (t)||\r _{st})-\ln Z={\rm tr}(\r \ln \r ) +\beta {\rm tr}(H\r 
)=\beta F
\end{equation}
Where $F$ is usual free energy. Therefore for the case of canonical 
distribution we have a slightly different form of H-theorem:
the free energy can only decrease if the system tends to equilibrium 
\cite{shlogl}\cite{stratthermo}\cite{landau}\cite{77}.

Is the physically relevant generalization  entropy is defined uniquely? 
This important question was investigated 
in \cite{donald}. The author showed that (\ref{st10}, \ref{st11}) with 
some other mathematical conditions are sufficient 
for the determination of (\ref{st9}).

The conclusion is the following:  (\ref{5}) {\it is correct 
generalization of  entropy to the more general
case, and a generalized H-theorem can be proved with the assumptions 
about the evolution of the system only.} 

Can we generalize 
(\ref{st10}) without any restrictions? 
If the answer is yes, then we can prove 
with this result a more general relation. 
Let us assume in formula (\ref{st10}) that 
\begin{equation}
\label{st15}
\S =c\hat {C_1}+(1-c)\hat {C_2},
\end{equation}
where $\hat {C_1}$ is defined by kraussian representation $A_\mu =|\mu
\rangle \langle 0|,\ \ \langle \mu |\acute \mu \rangle =\delta _{\mu \acute
\mu },\ \ \langle 0|0\rangle =1,\ \ 0\leq c\leq 1$. 
In other words for any
operator $\rho $:  
 $\hat {C_1}\rho =|0\rangle \langle 0|$. 
Now from (\ref{st10}), (\ref{st11})
we get 
\begin{eqnarray}
\label{st16}
&     & S(\S \rho ||\S \sigma)=
S(c\hat {C_1}\rho +(1-c)\hat {C_2}\rho||c\hat {C_1}\sigma 
+(1-c)\hat {C_2}\sigma )\nonumber \\
&\leq & cS(\hat {C_1}\rho ||\hat {C_1}\sigma ) 
+ (1-c)S(\hat {C_2}\rho  || \hat {C_2}\sigma )
\leq (1-c)S(\rho ||\sigma).
\end{eqnarray}
We see that if $\S $ is represented in the form (\ref{st15}) the ordinary
Lindblad inequality can be strengthened.

Now we need some general results
from  theory of linear operators \cite{matrix}. Let two 
hermitian operators $A$ and $B$ have the
spectrums $a_1\leq ...\leq a_n$, $b_1\leq ...\leq b_n$. For the spectrum $%
c_1\leq ...\leq c_n$ of the operator $C=A+B$ we have 
\begin{equation}
\label{st17}
a_1+b_k\leq c_k\leq b_k+a_n,\ \ b_1+a_k\leq c_k\leq a_k+b_n.
\end{equation}
where $k=1,...,n$. If 
\begin{eqnarray}
\label{st18}
&     &\rho ^{\prime } =\S \rho 
= c\hat {C_1}\rho +(1-c)\hat {C_2}\rho \nonumber \\
& =   & c| 0\rangle   \langle 0 |+(1-c)\sigma ,
\end{eqnarray}
and $\rho _1^{\prime } \leq ...\leq \rho _n^{\prime } $, 
$\sigma _1\leq ...\leq
\sigma _n$ are the spectrums of $\rho ^{\prime } $, $\sigma $ then we 
have 
\begin{eqnarray}
\label{st19}
&   &\rho _{1}^{\prime } -c\leq 
\sigma _{1}(1-c)\leq \min ( \rho _{1} ^{\prime },
\rho _{n} ^{\prime } -c),\nonumber \\
&   & \max (\rho _{1}^{\prime } ,\rho _{k}^{\prime } -c)
\leq \sigma _{k}(1-c)\leq
\rho _{k}^{\prime } ,
\end{eqnarray}
where $k=2,...,n$. We define $c(\S ,\rho )$ as the minimal eigenvalue 
of $
\rho ^{\prime} $ and $c(\S )=\min _\rho c(\S ,\rho )$ 
where minimization
is taken by all density matrices for the fixed Hilbert space. With the well
known results of operator theory \cite{matrix} we can write 
\begin{equation}
\label{st20}
c(\S )=\min _\rho \min _{\langle \psi |\psi \rangle =1}\langle \psi |\S
 \rho |\psi \rangle ,
\end{equation}
where the second minimization is taken by all normal vectors in the Hilbert
space. For any density matrix $\rho $ we get to the formula (\ref{st15}) 
where c is defined in (\ref{st20}) and $\hat {C_2}$ is some general evolution
operator. Now from (\ref{st15},\ref{st16},\ref{st20}) 
we get the strengthened Lindblad inequality 
\begin{equation}
\label{st21}
(1-c)S(\rho _1||\rho _2)\geq S(\S \rho _1||\S \rho _2).
\end{equation}
The equations 
(\ref{st20}), (\ref{st21}) 
are our general results. Of course there are many
evolution operators $\S $ with $c(\S )=0$ but later we shall
show that our
results can be nontrivial because for some simple but physically important
case $c(\S )$ is nonzero. 
From (\ref{st20}), (\ref{st21}) 
we immediately get to the strengthened $H$-theorem which gives us some 
information about the speed of relative entropy
decrease. An analog of (\ref{st21}) exists also in classical information 
theory \cite{chisar}.
Eq. (\ref{st21}) can be also regarded as a bound for entropy production. 
This quantity is very important in nonequilibrium 
statistical mechanics \cite{shlogl}\cite{stratthermo}.

Now about application of this result to quantum data processing.

Quantum information theory is a new field with potential
applications for the conceptual foundation of quantum mechanics. It appears
to be the basis for a proper understanding of the emerging fields of quantum
computation, communication and cryptography (see \cite{shum1} for 
references). 
Quantum information
theory is concerned with quantum bits (qubits) rather than bits. Qubits can
exist in superposition or entanglement states with other qubits, a notion
completely inaccessible for classical mechanics. More general, quantum
information theory contains two distinct types of problem. The first type
describes transmission of classical information through a quantum channel
(the channel can be noisy or noiseless). In such a scheme bits are 
encoded as some
quantum states and only these states or their tensor products are 
transmitted.
In the second case arbitrary superposition of these states or entanglement
states are transmitted. In the first case the problems can be solved by
methods of classical information theory, but in the second case new physical
representations are needed.

Mutual information is the most important
ingredient of information theory. In classical theory this quantity was
introduced by C.Shannon \cite{chisar}. 
The mutual information between two ensembles of
random variables $X$, $Y$ (for example these ensembles can be input and
output for a noisy channel) 
\begin{equation}
\label{st1}
I(X,Y)=H(Y)-H(Y/X),
\end{equation}
is the decrease of the entropy of $X$ due to the knowledge about $Y$, and
conversely with interchanging $X$ and $Y$. Here $H(Y)$ and $H(Y/X)$ are
Shannon entropy and mutual entropy \cite{chisar}.
 
Mutual information in the quantum
case must take into account the specific character of the quantum
information as it is described above. The first reasonable definition of
this quantity was  introduced by S.Lloyd \cite{lloyd}, and independently by
B.Schumacher and M.P.Nielsen \cite{shum2}. 
Suppose a
quantum system with density matrix 
\begin{equation}
\label{st2}
\rho =\sum_ip_i|\psi _i\rangle \langle \psi _i|,\ \ \sum_ip_i=1.
\end{equation}
We only assume that $\langle \psi _i|\psi _i\rangle =1$ and the states may
be nonorthogonal. The noisy quantum channel can be described by some general
quantum evolution operator $\S $.

As follows from the definition of quantum information transmission, a
possible distortion of entanglement of $\rho $ must be taken into account.
In other words a definition of mutual quantum information must contain the
possible distortion of the relative phases of the quantum ensemble $\{|\psi
_i\rangle \}$. Mutual quantum information is defined as 
\cite{lloyd}\cite{shum2}
\begin{equation}
\label{st4}
I(\rho ;\S )=S(\S \rho )-S(\hat 1^R\otimes \S (|\psi ^R\rangle
\langle \psi ^R|)),
\end{equation}
\begin{equation}
\label{st5}
\hat 1^R\otimes \S (|\psi ^R\rangle \langle \psi ^R|)=\sum_{i,j}\sqrt{%
p_ip_j}|\phi _i^R\rangle \langle \phi _j^R|\otimes \S (|\psi _i\rangle
\langle \psi _j|).
\end{equation}
Where $S(\rho )$ is the entropy of von Newman and $\psi ^R$ is a
purification of $\rho $ 
\begin{equation}
\label{st6}
|\psi ^R\rangle =\sum_i\sqrt{p_i}|\psi _i\rangle \otimes |\phi _i^R\rangle
,\ \ \langle \phi _j^R|\phi _i^R\rangle =\delta _{ij},
\end{equation}
\begin{equation}
\label{st7}
{\rm tr}_R|\psi ^R\rangle \langle \psi ^R|=\rho ,
\end{equation}
here $\{|\phi _i^R\rangle \}$ is some orthonormal set. The definition is
independent of the concrete choice of this set \cite{shum1}. 
Mutual quantum
information is the decrease of  entropy after the action of $\S $ due to
the possible distortion of entanglement state. This quantity is not
symmetric with respect to the interchanging of  input and  output and can be
positive, negative or zero in contrast with Shannon mutual information
in classical theory.
It has been shown that (\ref{st4}) can be the upper bound of
the capacity of a quantum channel 
\cite{alah}. 
Using this value the authors \cite{alah}
have been proved the converse coding theorem for a quantum source with 
respect
to the so called entanglement fidelity \cite{shum1}. 
This fidelity is absolutely adequate for quantum
data transmission or compression.

In the \cite{shum2} the authors prove a data
processing inequality 
\begin{equation}
\label{st8}
I(\rho ;\S _1 )\geq I(\rho ;\S _2 \S _1 ).
\end{equation}
The quantum information can not increase after action of $\S$.
In \cite{alah} we found an alternative derivation 
of this result which is
simpler than the derivation of \cite{shum2}. 
In the present paper we show that this inequality can
be strengthened. The data processing inequality is a very important 
property of
mutual information. This is an effective tool for proving general results
and the first step toward identification of a physical quantity as mutual
information.
 
Now we briefly recall the derivation of the data processing
inequality.
The formalism of relative quantum entropy is
very useful in this context.
We have 
\begin{eqnarray}
\label{st12}
&   & S(\hat{1}^R\otimes \S (|\psi ^R\rangle \langle \psi ^R|)||\hat
{1}^R\otimes \S (\rho ^R\otimes \rho  ))\nonumber \\
& =&-S(\hat{1}^R \otimes \S (|\psi
^R\rangle \langle \psi ^R|))+S(\rho ^R)+S(\S \rho )).
\end{eqnarray}
Here 
\begin{equation}
\label{st13}
\rho ^R=\sum_{i,j}\sqrt{p_ip_j}|\phi _i^R\rangle \langle \phi _j^R|\langle
\psi _i|\psi _j\rangle .
\end{equation}
Now from the Lindblad inequality we have 
\begin{eqnarray}
\label{st14}
&      & S(\hat {1^R}\otimes \S (|\psi ^R\rangle \langle \psi ^R|)||\hat
{1^R}\otimes \S (\rho ^R\otimes \rho ))\nonumber \\
& \geq & S(\hat {1^R}\otimes \hat
{S_1}\S _2 (|\psi ^R\rangle \langle \psi ^R|)||\hat {1^R}\otimes \S _1 \S 
_2 (\rho ^R\otimes \rho )).
\end{eqnarray}
From this formula we have (\ref{st8}).
 
Now we can prove the strengthened data processing inequality. 
Let in (\ref{st14}) $%
\S _2 $ is represented in the form (\ref{st15}). 
From (\ref{st10},\ref{st15}) we get 
\begin{eqnarray}
\label{st22}
&   & S(\hat{1}^R\otimes \S _2\S _1(|\psi ^R\rangle \langle \psi ^R|)
||\hat {1}^R\otimes \S _2\S _1(\rho ^R\otimes \rho  ))\nonumber \\
& \leq &-(1-c)S(\hat{1}^R \otimes \hat C_2\S _1(|\psi
^R\rangle \langle \psi ^R|))+S(\rho ^R)+(1-c)S(\hat C_2\S \rho )).
\end{eqnarray}
And we have 
\begin{equation}
\label{st23}
(1-c(\S _2 ))I(\rho ;\S _1)\geq I(\rho ;\S _2\S _1).
\end{equation}
Now we consider the simplest example of noisy
quantum channel: Two dimensional, two- Pauli channel \cite{dublepauli} 
with the following Krauss representation
\begin{equation}
\label{st24}
A_1=\sqrt{x}\hat 1,\ \ A_2=\sqrt{(1-x)/2}\sigma _1,\ \ A_3=-i\sqrt{(1-x)/2}%
\sigma _2,\ \ 0\leq x\leq 1,
\end{equation}
where $\hat 1$, $\sigma _1$, $\sigma _2$ are the unit matrix and the first
and the second Pauli matrices. 
The (\ref{st24}) also has physical meaning as an evolution operator for a
two-dimensional open system.
Any density matrix in two-dimensional Hilbert
space can be represented in the Bloch form 
\begin{equation}
\label{st25}
\rho =(1+\vec a\vec \sigma )/2,
\end{equation}
where $\vec a$ is a real vector with $|\vec a|\leq 1$. Now we have 
\begin{equation}
\label{st26}
\S _{TP} ((1+\vec a\vec \sigma )/2)=(1+\vec b\vec \sigma )/2,
\end{equation}
where $\vec b=(a_1x,a_2x,a_3(2x-1))$. After simple calculations we get 
\begin{equation}
c(\S _{TP} )=(1-|2x-1|)/2.
\end{equation}
We conclude by reiterating the main results:
the Lindblad inequality can be generalized.
We have presented results not only about increasing of  entropy
and decreasing of  mutual quantum information
but also about the speed of these processes.




\references
\bibitem{shlogl}F.Shlogl, Phys. Rep.,{\bf 62}, 268, (1980).
  
 \bibitem{stratinfo} R.L. Stratanovich, {\it Information Theory}, 1975, 
Moscow, Nauka.   

\bibitem{stratthermo} R.L. Stratanovich, {\it 
Nonequilibriun, Nonlinear Thermodynamics}, 1985, Moscow, Nauka.   

\bibitem{landau} L.D. Landau, and E.M. Lifshits, {Statistical Physics}, 
(Moscow, Nauka, 1976).

\bibitem{pop}R.P. Poplavsky, {\it Thermodynamics of
information processes}, 1981, Moscow, Nauka.

\bibitem{77} The literature on this particular point is too broad to cite 
every work. A recent book of M.C. Mackey,
{\it Time's Arrow: The Origins of Thermodynamic Behavior}, 
(Springer-Verlag, Berlin, 1992) provides a good review.
See also: M.C. Mackey, Rev.Mod.Phys., {\bf 61}, 981, (1989); Yu.L. 
Klimontovich, Uspekhi Fizicheskikh Nauk, {\bf 164},
811, (1994); Yu.L. Klimontovich, {\it Statistical Physics}, (Moscow, 
Nauka, 1982).

\bibitem{wehrl}A. Wehrl, Rev.Mod.Phys., {\bf 50}, 221, (1978).

\bibitem{donald} M.J. Donald, Comm.Math.Phys., {\bf 105}, 13, (1986).

\bibitem{open} For a comprehensive review and further references about 
nonequilibrium steady states
in open systems, see:
                     B.~Schmittmann and R.K.P.~Zia,
		in {\it Phase Transitions and Critical Phenomena},
		eds. C.~Domb and J.L.~Lebowitz
		(Academic Press, New York, 1996).


\bibitem{barnet}S.M. Barnett and S.J.D. Phoenix,
Phys. Rev. A {\bf 44}, 535, (1991).

\bibitem{shum1}  B.Schumacher, 
Phys. Rev. A {\bf 54}, 2614, (1996).

\bibitem{shum2}  B.Schumacher and M.A. Nielsen, Phys. Rev. A {\bf 54}, 2629,
(1996).

\bibitem{alah}  A.E. Allahverdyan and D.B. Saakian, eprint quant-ph/9702023.

\bibitem{lloyd}  S. Lloyd, Phys. Rev. A {\bf 55}, R1613-1622,
(1997); also e-print quant-ph/9604015.

\bibitem{krauss}  K.Kraus, Ann. Phys., {\bf 64}, 311, (1971).

\bibitem{ohya}  M.Ohya, Rep. Math. Phys., {\bf 27}, 19, (1989).

\bibitem{umegaki}  H.Umegaki, Kodai Math. Sem. Rep., {\bf 14}, 59, (1962).

\bibitem{lindblad}  G.Lindblad, Commun. Math. Phys., {\bf 40}, 147, (1975).

\bibitem{lax} M. Lax, Phys.Rev., {\bf 145}, 110, (1966).

\bibitem{knight} S.J.D. Phoenix and P.L. Knight, Ann. Phys.(N.Y) {\bf 186},
381, (1988). 

\bibitem{dublepauli}  C.H. Bennett, 
C.A. Fuchs, J.A. Smolin, eprint quant-ph/9611006.

\bibitem{chisar}  I.Csiszar, J.Korner, {\it Information Theory}, 
(Akademiai Kiado, Budapest,1981).

\bibitem{matrix}  F.Gantmacher, {\it The Theory of matrices},
Moscow, Nauka 1983.
\end{document}